\documentclass[twocolumn]{aastex631}

\usepackage{orcidlink} 
\usepackage[T1]{fontenc}
\usepackage{ae,aecompl}

\usepackage{graphicx}	
\usepackage{amsmath}	
\usepackage{amssymb}	
\usepackage{natbib}

\usepackage{hyperref}
\hypersetup{
	colorlinks=true,
	urlcolor=blue,    
	linkcolor=black,  
	citecolor=blue,   
}

\defcitealias{Pittordis_2018}{PS18}
\defcitealias{Pittordis_2019}{PS19}
\defcitealias{Badry_2021}{ERH}
\defcitealias{Banik_2018_Centauri}{BZ18}

\hypersetup{citecolor=blue, 
            linkcolor=red, 
            menucolor=blue, 
            urlcolor=blue}  

\newcommand{\cm}{{~\rm cm}}
\newcommand{\m}{{~\rm m}}
\newcommand{\km}{{~\rm km}}
\newcommand{\s}{{~\rm s}}
\newcommand{\h}{{~\rm h}}

\newcommand{\g}{{~\rm g}}
\newcommand{\G}{{~\rm G}}
\newcommand{\K}{{~\rm K}}
\newcommand{\erg}{{~\rm erg}}

\begin{document}

\title{Identifying a point-symmetric morphology in supernova remnant Cassiopeia A: explosion by jittering jets}


\author{Ealeal Bear}\author{Noam Soker\,\orcidlink{0000-0003-0375-8987}}
\affiliation{Department of Physics, Technion, Haifa, 3200003, Israel; ealeal44@technion.ac.il; soker@technion.ac.il}


\begin{abstract}
We identify a point-symmetric morphology of the supernova remnant (SNR) Cassiopeia A compatible with shaping by at least two, and more likely more than four, pairs of opposite jets, as expected in the jittering jets explosion mechanism (JJEM) of core-collapse supernovae. Using an old Spitzer Telescope infrared map of argon, we identify seven pairs of opposite morphological features that we connect with lines that cross each other at the same point on the plane of the sky. The opposite morphological features include protrusions, clumps, filaments, and funnels in the main SNR shell. In addition to these seven symmetry axes, we find two tentative symmetry axes (lines). These lines form a point-symmetric wind-rose. We place this point-symmetric wind-rose on a new JWST and X-ray images of Cassiopeia A. We find other morphological features and one more symmetry axis that strengthen the identified point-symmetric morphology. Not all symmetry axes correspond to jets; e.g., some clumps are formed by the compression of ejecta between two jet-inflated lobes (bubbles). The robust point-symmetric morphology in the iconic Cassiopeia A SNR strongly supports the JJEM and poses a severe challenge to the neutrino-driven explosion mechanism. 
\end{abstract}


\section{Introduction} 
\label{sec:intro}

The collapse of the inner core of a massive star to form a neutron star (NS) releases $\simeq {\rm few} \times 10^{53} \erg$ of gravitational energy, most of which is carried by neutrinos (e.g., \citealt{Janka2012}). The rest explodes the star in a core-collapse supernova (CCSN) having explosion energy of $E_{\rm exp} \approx 10^{49} -10^{52} \erg$ (e.g., \citealt{Burrows2013}). There is no consensus on the CCSN explosion mechanism. Recent studies consider either the delayed neutrino explosion mechanism that is based on a small fraction of the energy that the neutrinos carry (e.g., \citealt{Mulleretal2019Jittering, Fujibayashietal2021, Fryeretal2022, Bocciolietal2022, Nakamuraetal2022, Olejaketal2022, Bocciolietal2023, Burrowsetal2023, BoccioliRoberti2024}; for some earlier studies see, e.g., \citealt{BetheWilson1985, Hegeretal2003, Janka2012, Nordhausetal2012}) or the jittering jets explosion mechanism (JJEM) that is based on the accretion process of mass onto the newly born NS that launches jittering jets (e.g., \citealt{Soker2018B, Soker2020RAA, Soker2022SNR0540, Soker2022Boosting, Soker2022Rev, Soker2023gap, Quataertetal2019,  ShishkinSoker2021, ShishkinSoker2022, ShishkinSoker2023, AntoniQuataert2022, AntoniQuataert2023, WangShishkinSoker2024}; for some earlier studies see, e.g.,  \citealt{Soker2010, PapishSoker2011, GilkisSoker2014, GilkisSoker2016}).  

The JJEM expectation is that in some, but not all,  CCSN remnants (CCSNRs), there will be observable signatures of two or more pairs of jets as point-symmetrical morphologies. 
Neither interaction with the interstellar medium nor instabilities can explain point symmetry (e.g., \citealt{SokerShishkin2024}, and a review by \citealt{Soker2024Rev}). Instabilities exist in the delayed neutrino explosion mechanism (e.g., \citealt{Wongwathanaratetal2015, Wongwathanaratetal2017, BurrowsVartanyan2021, Vartanyanetal2022, Orlando2023}) and are also expected in the JJEM. They mainly smear the point-symmetric morphological signatures and introduce more stochastic structural features.  

In the last year, the JJEM received strong support for identifying point-symmetrical structures in several CCSNRs, {{{{ similar in many aspects to morphologies of some planetary nebulae (e.g., \citealt{Soker2024PN}). }}}} These include the Vela SNR (\citealt{Soker2023SNRclass, SokerShishkin2024}; for new observations, see \citealt{Mayeretal2023}), SNR N63A \citep{Soker2024N63A}, SNR G321.3–3.9 (\citealt{ShishkinSoker2024}; for new observations, see \citealt{Mantovaninietal2024}), and tentative identifications in the SNR CTB 1 \citep{BearSoker2023RNAAS}, in SNR 1987A \citep{Soker2024NA1987A} and in SNR G107.7-5.1 \citep{Soker2024CFs}. There is also a point-symmetrical structure in the velocity map of SNR 0540-69.3 \citep{Soker2022SNR0540}. Identifying a point-symmetrical structure is not always trivial and might require combining images in two or more emission bands, like radio and X-ray; this is the case with identifying the point symmetry in SNR G321.3-3.9 \citep{ShishkinSoker2024}. 
 
The reasons for the difficult identification of point symmetry in CCSNRs are as follows. Two or more pairs of opposite jets with their symmetry axes in different directions shape the ejecta to have a point-symmetrical structure. However, out of several to a few tens of jet-launching episodes in the JJEM, only the last jets will leave an observable point-symmetrical structure. The first jets are choked well inside the star, and the structure they shape is smeared by instabilities, the hot ejecta with a high sound speed, and the large-scale asymmetrical mass ejection that also imparts a natal kick velocity to the NS. The large-scale asymmetry of the ejecta might cause the NS natal kick, not the jets, e.g., the kick is not along the main-jet axis (e.g., \citealt{BearSoker2023RNAAS}). Namely, in the JJEM, the NS natal kick mechanism might be the same as that of the neutrino-driven explosion mechanism. However, it might be that the process that imparts a kick velocity to the NS is part of the JJEM . A very early asymmetrical jet pair or two causes the NS kick velocity, as was suggested recently by \cite{BearShishkinSoker2024}, who termed this process kick by early asymmetrical pairs (Kick-BEAP) mechanism. In \cite{BearSoker2018}, we studied several CCSNRs (updated in \citealt{BearSoker2023RNAAS}),  showed that the main-jet axis tends to avoid small angles to the natal-kick velocity, and explained that the JJEM is compatible with this finding. In Cassiopeia A we found the projection of this angle to be $88^\circ$. 
 \cite{YamasakiFoglizzo2008} consider the possibility that the NS kick velocity results from the $l=1$ instability modes of the spiral standing accretion shock instability and that this leads to a kick velocity that tends to be perpendicular to the spin axis of the NS. In the JJEM, the misalignment of the kick velocity is with the main-jet axis, which in some cases is aligned with the NS spin axis and in some cases not.

By the hot ejecta, we refer to the thermal heat of the entire ejecta from the explosion itself, the heat deposited by radioactive isotopes in specific locations termed nickel-bubbles, and the heating by the reverse shock (which can be complicated, e.g., \citealt{HwangLaming2012}). Nickel bubbles can form dense knots, as observed in Cassiopeia A (e.g., 
\citealt{Fesen2001, HammellFesen2008, MilisavljevicFesen2013, FesenMilisavljevic2016}). However, we note that bubbles alone will not form point-symmetric morphological features. More likely, different bubbles form different structures of knots, and therefore, nickel bubbles tend to smear point-symmetric morphological features.  

If only one pair of jets leaves an observable imprint on the CCSNR, then there is one axis of symmetry. An axial symmetry might reveal itself by a barrel-shaped morphology (e.g., \citealt{BearSoker2017, Akashietal2018, Soker2023SNRclass}) or one pair of two opposite protrusions that are termed `ears' (e.g., \citealt{Bearetal2017, GrichenerSoker2017ears, Soker2023SNRclass}). An ear is defined as a protrusion that has a base smaller than the main diameter of the nebula/SNR, a cross-section that decreases with distance from the center, and an intensity, color, and/or boundary that differ from those of the main nebula/SNR.   

It is important to consider that in the JJEM, the energy in each jet $E_{\rm jet}$ is a small fraction of the total explosion energy, i.e.,  $E_{\rm jet} \approx 0.01 E_{\rm exp} - 0.1 E_{\rm exp}$, explaining the finding that the energy required to inflate an ear in CCSNRs is 
$\approx 1-10\%$ the total kinetic energy of the SN ejecta \citep{GrichenerSoker2017ears}. This largely differs from the jet's explosion energy in the magneto-hydrodynamical jet model (e.g., \citealt{Khokhlovetal1999, Maedaetal2012, LopezCamaraetal2013, BrombergTchekhovskoy2016,  Nishimuraetal2017, WangWangDai2019RAA, Grimmettetal2021, Gottliebetal2022, ObergaulingerReichert2023, Urrutiaetal2023a}) where a rapidly rotating pre-collapse core leads to only two opposite jets, each carry about half of the explosion energy. In any case, the magneto-hydrodynamical jet model can operate only in rare cases of rapidly rotating pre-collapse cores. 

 In the JJEM, intermittent accretion disks and accretion belts, i.e., sub-Keplerian accretion inflow that leaves open funnels along the poles (e.g., \citealt{SchreierSoker2016}), launch the jet via magneto-hydrodynamical processes. Magnetic fields play crucial roles in the launching process of jets in the JJEM (e.g., \citealt{Soker2019RAA}). Neutrino heating boosts the jet's energy after launching \citep{Soker2022Boosting}. 
Other properties of the jittering jets include $\approx {\rm several}$ to ${\rm few} \times 10$ pairs of jittering jets, each pair carrying a total mass of $\approx 10^{-3} M_\odot$ and having a typical sub-relativistic velocity of $v_{\rm j} \simeq 10^5 \km \s^{-1}$ (\citealt{Guettaetal2020} used neutrino observations to constrain jets in most CCSNe to be non-relativistic). 

In this study, we identify a point symmetric morphology in an argon-velocity map from \cite{DeLaneyetal2010} of the well-studied SNR Cassiopeia A (section \ref{sec:Argon}) and explain it in the frame of the JJEM (section \ref{sec:Summary}). 
We are partly motivated by the new JWST images of Cassiopeia A \citep{Milisavljevicetal2024} that present the two eastern protrusions, called ears, in a new light. 
\cite{Soker2017RAA} already argued that the JJEM better explains the morphology of Cassiopeia A than the delayed neutrino explosion mechanism, but he did not identify a point symmetry. \cite{PapishSoker2014Planar}  speculated that planar jittering-jets are behind the morphology of the Cassiopeia A supernova remnant. In this study, we confirm their speculation. 

Although the structure of Cassiopeia A in three-dimension is well studied and the main two opposite jets are well known (e.g., \citealt{Reedetal1995, LamingHwang2003, Hinesetal2004, Hwangetal2004, Fesenetal2006, MilisavljevicFesen2013,  Zhanetal2022, Ikedaetal2022, Kooetal2023, Sakaietal2023,  Milisavljevicetal2024}), this is the first study that identifies a point symmetry and relates it to the explosion mechanism itself. We take the jets to have exploded Cassiopeia A in the frame of the JJEM, rather than being a post-explosion process, as, e.g., \cite{Orlandoetal2021} suggested for the two main opposite jets of Cassiopeia A. 
We then relate the identified point-symmetrical structure in the argon velocity map to maps in other electromagnetic bands (section \ref{sec:OtherMaps}). We summarize this study in section \ref{sec:Summary}.

\section{The point symmetry in the Argon map} 
\label{sec:Argon}

Cassiopeia A is a clumpy SNR and has a typical CCSN explosion energy of $E_{\rm exp} \simeq 10^{51} \erg$ (e.g., \citealt{Willingaleetal2003}). \cite{Willingaleetal2003} consider the distribution of the prominent clumps either in an equatorial plane or in the polar regions of an axially symmetrical structure. They comment on the mass concentration within $\pm 30^\circ$ of the equatorial plane. Following later studies, e.g., \cite{MilisavljevicFesen2013}, we take the optically emitting material mass distribution to be in and around an equatorial plane. 
 The optically emitting material is not the entire ejecta of Cassiopeia A, as the analysis of X-ray observations show (e.g., \citealt{HwangLaming2012}). For the present morphological study, the dense ejecta in an equatorial plane that we consider here signifies the explosion morphology even if less dense ejecta expand along the two polar directions. 

\begin{figure}
\begin{center}
\includegraphics[trim=10.6cm 2.1cm 0.0cm 2.7cm ,clip, scale=0.52]{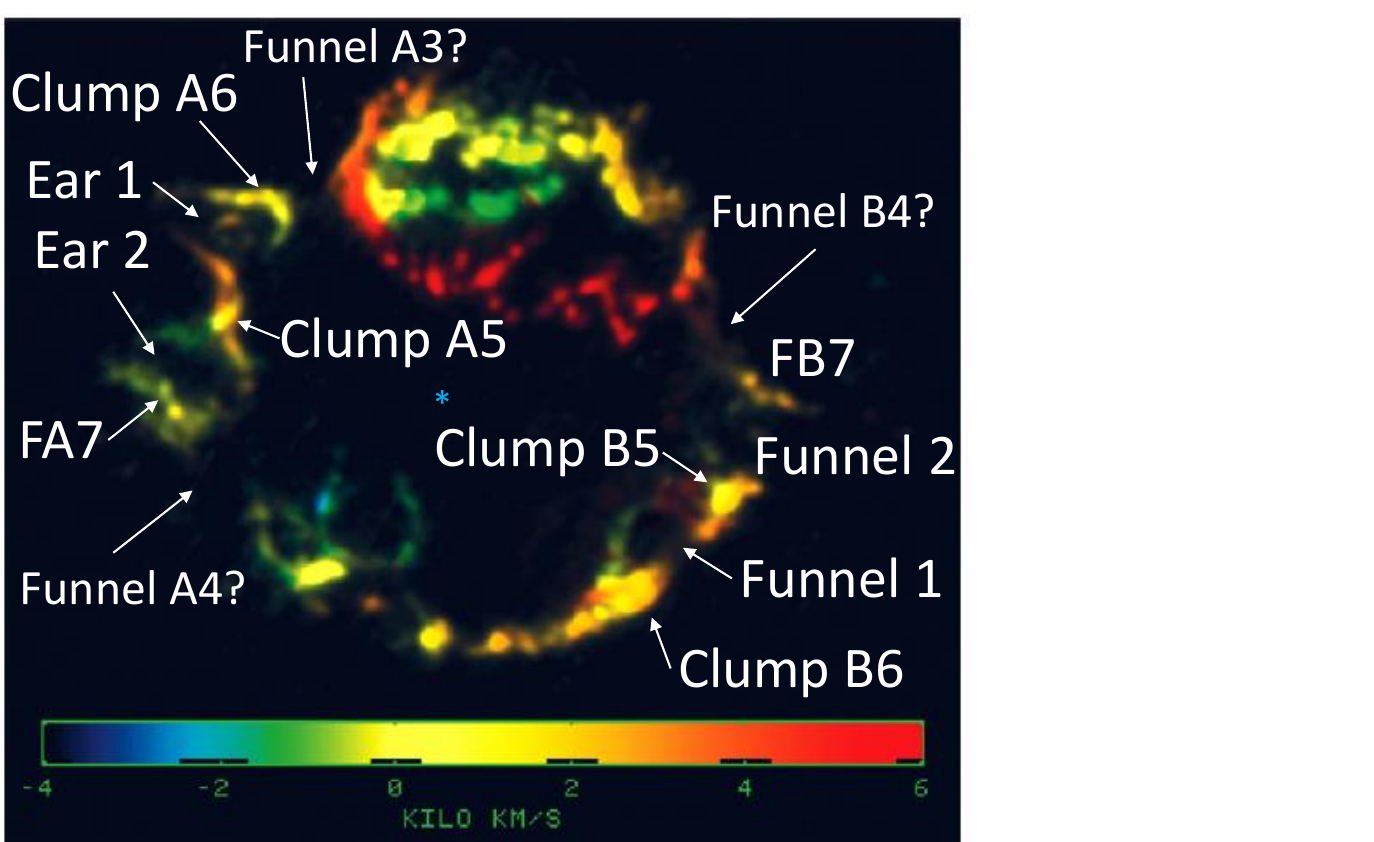}
\includegraphics[trim=10.6cm 2.1cm 0.0cm 2.7cm ,clip, scale=0.52]{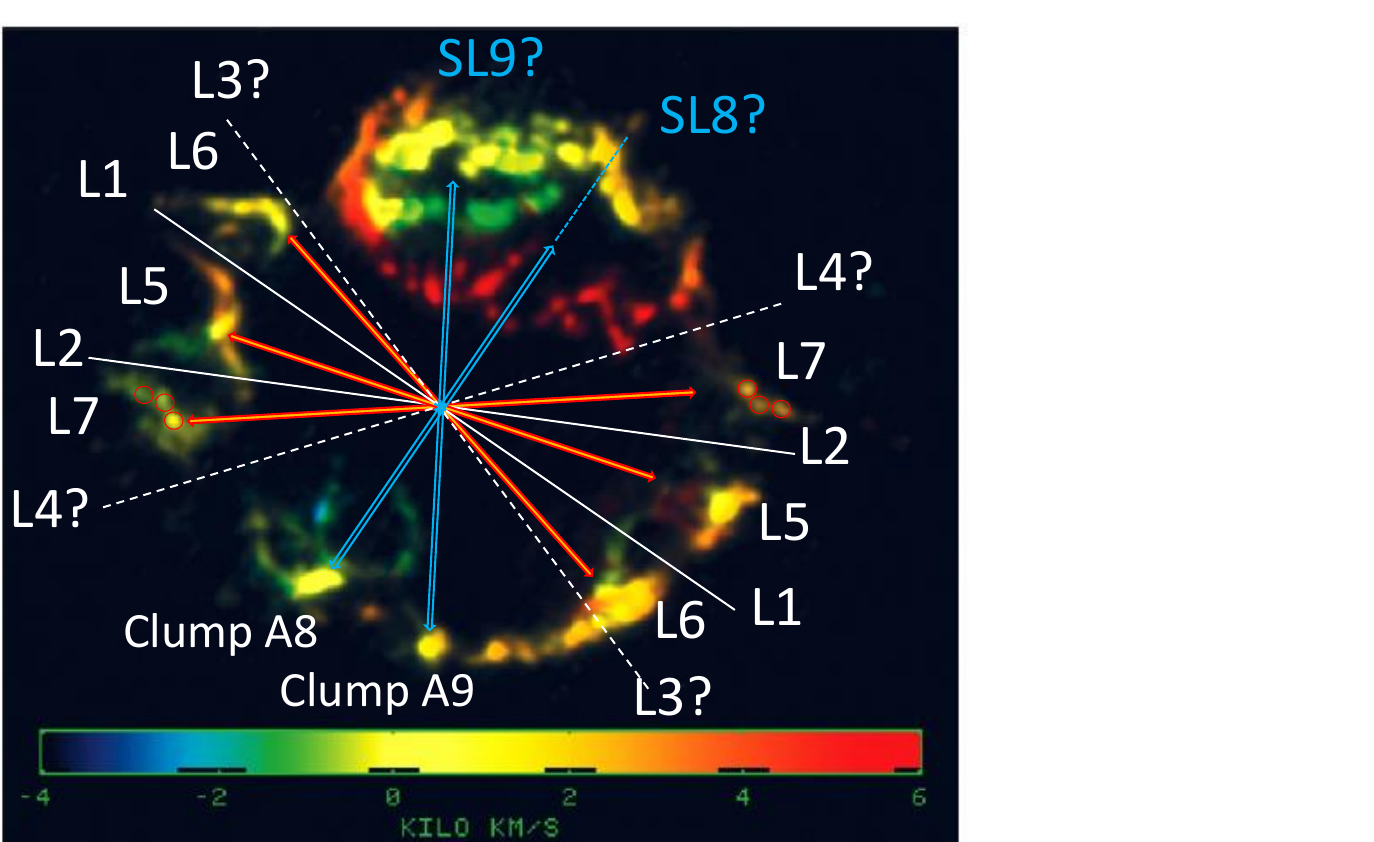}
\end{center}
\caption{A velocity map (by Doppler shift) of Cassiopeia A in the $6.99 \mu$m [Ar \textsc{ii}] line taken from \cite{DeLaneyetal2010}, with the addition of our labeling (the two panels have different labeling). The velocity scale is from $-4000 \km \s^{-1}$ (deep blue) to $6000 \km \s^{-1}$ (deep red). We mark ears, funnels, clumps, and two filaments FA7 and FB7. The three yellow-filled red (seem orange) double-headed arrows point at opposite clumps. The solid-white lines are along two jets' axes that we suggest shaped the ears and opposite funnels. From this map alone, we are less confident with symmetry axes L3 and L4 being real jets' axes. We further draw two speculative symmetry axes, SL8 and SL9. Line SL8 goes from the bright isolated Clump A8 through the center and is marked by the double-headed light-blue arrow; the arrow's center is at the cross point of the other symmetry lines. Note the faint clump on the continuation of the SL8 line at the northern tip of the dashed-light-blue line. Similarly, speculative line SL9 starts from Clump A9 and goes through the cross point of the other lines.  All these lines form the \textit{point-symmetric wind-rose }of Cassiopeia A.   
}
\label{Fig:ArgonMaps1} 
\end{figure}

Figure \ref{Fig:ArgonMaps1} presents an argon velocity map from \cite{DeLaneyetal2010}.
In the upper panel, we added labeling of morphological features relevant to our study, while in the lower panel, we connected opposite morphological features that we identified to reveal Cassiopeia A's point symmetry. 
We mainly refer to yellow-colored features, implying a close to zero line-of-sight velocity. These regions are more or less in the plane of the sky (e.g., \citealt{Willingaleetal2003, MilisavljevicFesen2013}).  
The light-blue asterisk marks the center of the four lines L1-L4 and the center of the three double-headed arrows L5-L7. We do not claim this is the position of the explosion itself; we identify the best cross point of the different symmetry lines. 

 Before further analyzing the argon map, we mention the selection effects that the emission line by the singly ionized argon introduces. Photoionization-recombination equilibrium calculations by \cite{LamingTemim2020} show that the relevant singly ionized elements (in the present case Ar$^{+}$) form in the unshocked ejecta at densities of $10 - 100 {~\rm amu} \cm^{-3}$, which are one to two orders of magnitude higher than the average density of unshocked ejecta \citep{HwangLaming2012}. Therefore, the infrared emission of $6.99 \mu$m [Ar \textsc{ii}], hence the image of figure \ref{Fig:ArgonMaps1}, results from the cooling zones of the reverse shock where the reverse shock slows down inside dense clumps. Figure \ref{Fig:ArgonMaps1}, which emphasizes dense and slow clumps, serves our goal of revealing the point-symmetric structure of Cassiopeia A. We use Figure \ref{Fig:ArgonMaps1} to build the point-symmetric wind-rose, which we extend and make robust with newer less biased observations in Section \ref{sec:OtherMaps}   

We assume that each of the two ears was inflated by a jet. If a jet breaks out from an ear, it leaves behind a funnel (or a nozzle or a chimney). Since one jet in a pair can be more energetic than its counter jet \citep{Soker2024N63A}, the counter jets to those that inflated the two ears on the east may brook out in the west. We, therefore, pair Ear 1 and Ear 2 with Funnel 1 and Funnel 2, respectively. These are lines L1 and L2 in the lower panel of Figure \ref{Fig:ArgonMaps1}. 
We are less confident about the pair of Funnel A4 with Funnel B4, and cannot find an actual counter funnel to Funnel A3 in the Argon velocity map. We, therefore, draw the symmetry lines (axes) L3 and L4 with dashed lines. We return to line L3 in section \ref{sec:OtherMaps}.

 A property of jet-inflated structures, such as cavities, lobes, ears, and nozzles, is that they do not necessarily move faster than the rest of the ejecta. The fast jets inflate these structures, but the high pressure of the jet-ejecta interaction region might push some gas segments backward (towards the center) and sideways. Only the front of the interaction region moves faster; in some cases, it expands to large distances, and its density becomes too low to be observed. The opening the jet has left in the interaction zone becomes a nozzle (e.g., \citealt{Soker2024Keyhole}). 
  
The double-headed yellow-filled red arrows connect opposite isolated bright clumps. The double-headed arrow L7 connects two bright clumps opposite each other at the base of the respective filaments, FA7 and FB7. Again, exploding jets in the JJEM are expected to be unequal in power; therefore, opposite clumps may be at a different distance from the center. We draw three circles around the three bright clumps that compose filament FB7 (three red circles on the west side of the bottom panel of Figure \ref{Fig:ArgonMaps1}). We found that if we rotate this structure of three circles by $170^\circ$ around itself counterclockwise and place it on the opposite side, it matches the three bright clumps of counter filament FA7. This further strengthens our identification of a point-symmetrical morphology in Cassiopeia A and our pairing of Ear 2 with Funnel 2.    

Despite the very complicated structure of Cassiopeia A, the argon morphology of SNR Cassiopeia A reveals the following prominent point-symmetry properties. 
\begin{enumerate}
    \item There are three clear pairs of bright opposite argon clumps, connected by the double-headed yellow-filled red arrows in the lower panel of Figure \ref{Fig:ArgonMaps1}. 
    \item The two ears on the eastern side suggest two separate jets. The empty zones in the shell on the opposite (western) side, Funnel 1 and Funnel 2 suggest pairs of opposite jets. This is signified by the opposite pairs of three small clumps that compose filaments FA7 and FB7 (marked by three red circles in Figure \ref{Fig:ArgonMaps1}). 
    \item There are hints for the operation of two other pairs of jets, marked by the dashed-white lines in the lower panel of  Figure \ref{Fig:ArgonMaps1}. 
    \item These four pairs of jets are in one plane. We return to this property, which is expected in the JJEM, in section \ref{sec:Summary}. 
    \item Opposite structures do not have the same distance from the center nor the same size. The explanation (worked out by \citealt{Soker2024N63A}) is that the intermittent accretion disk around the newly born NS, which launches the jets, has no time to fully relax. Therefore, the two sides of the accretion disk are likely to be unequal. Therefore, the two opposite structural features that the two unequal jets shape might differ in shape, distance from the center of the explosion, and size. 
    Another departure from pure-point symmetry, which is not so rare in planetary nebulae (e.g., \citealt{SokerHadar2002} who estimated that $\approx 10 \%$ of planetary nebulae have bent morphology) and in jet-inflated bubbles in cluster cooling flows (see \citealt{Soker2024CFs} for relation to CCSNe), is that two opposite structures might not be precisely at $180^\circ$ to each other; this is termed bent-symmetry. The structure of three circles of the bright knots of filament FB7 is rotated around itself by $170^\circ$ counterclockwise to match filament FA7. \cite{Picquenotetal2021} argue that because opposite red-shifted and blue-shifted features in Cassiopeia A are not exactly at $180^\circ$ to each other,  jets did not drive the explosion of Cassiopeia A. We strongly disagree with this claim.   
\end{enumerate}

Overall, we find that the eastern-western elongation of Cassiopeia A was not curved by two opposite wide jets as some studies suggested in the past, e.g., \cite{MilisavljevicFesen2013} who inferred an opening half-angle of $\approx 40^\circ$, but instead composed of two and likely four and more pairs of jets. The total angle from line L3 to L4 is $70^\circ$, similar to the opening whole angle of $\approx 80^\circ$ that \cite{MilisavljevicFesen2013} measured. This wide-angle outflow comprises at least four pairs of jets (there might be weaker jets that we do not identify).

\section{Point symmetry in other wavelengths} 
\label{sec:OtherMaps}

This study was motivated in part by new studies of Cassiopeia A in the infrared (IR) by \cite{Milisavljevicetal2024} and in X-ray by \cite{Vinketal2024}. \cite{Milisavljevicetal2024} argue that the structure of Cassiopeia A is compatible with the delayed neutrino explosion mechanism and radioactive heating. Radioactive heating after the explosion in Cassiopeia A, see, e.g., \cite{MilisavljevicFesen2015}, also plays a role in the JJEM.   

\begin{figure}
\begin{center}
\includegraphics[trim=5.7cm 3.3cm 1.9cm 0.3cm ,clip, scale=0.43]{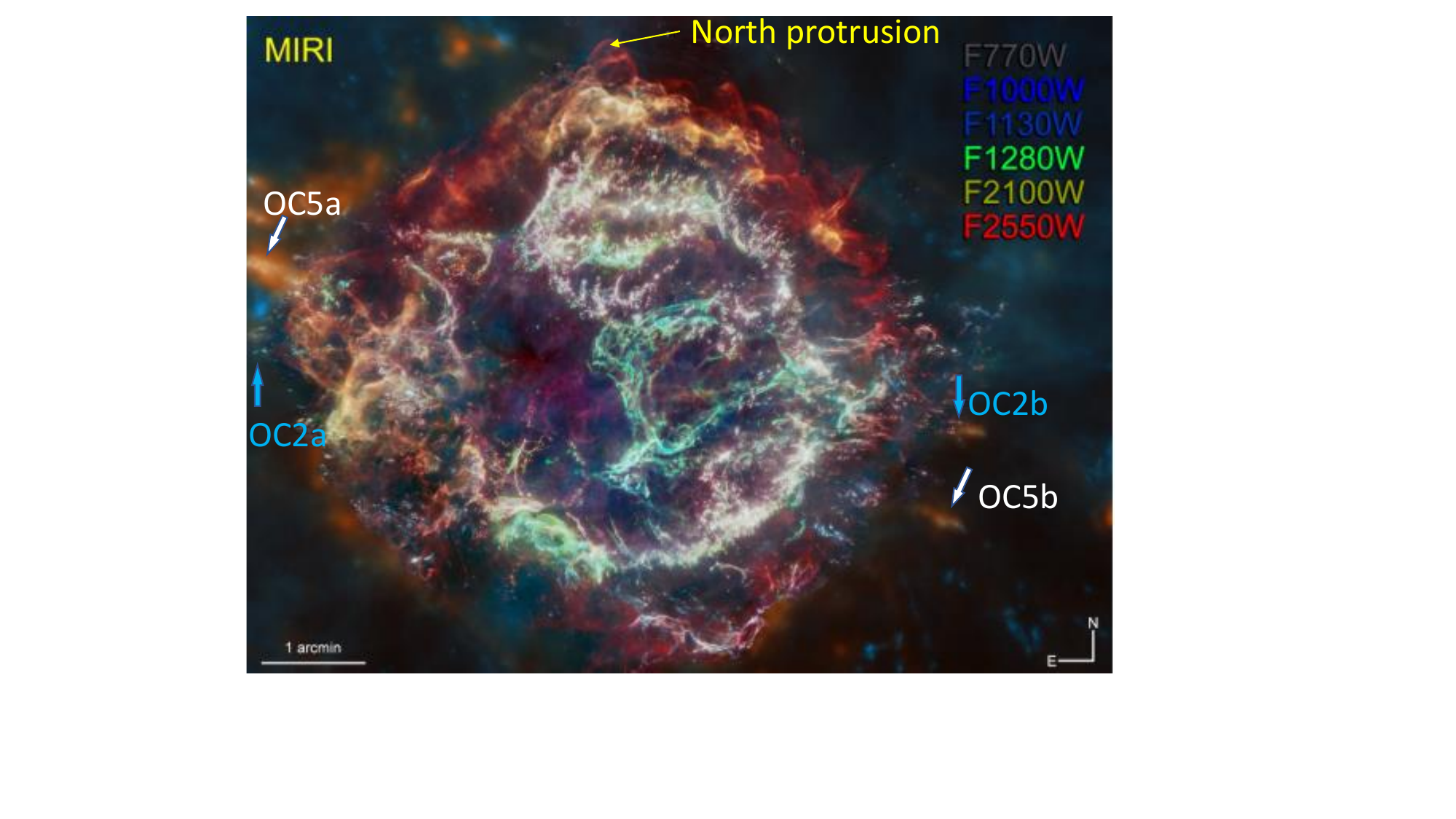}
\includegraphics[trim=5.7cm 3.3cm 1.9cm 0.3cm ,clip, scale=0.43]{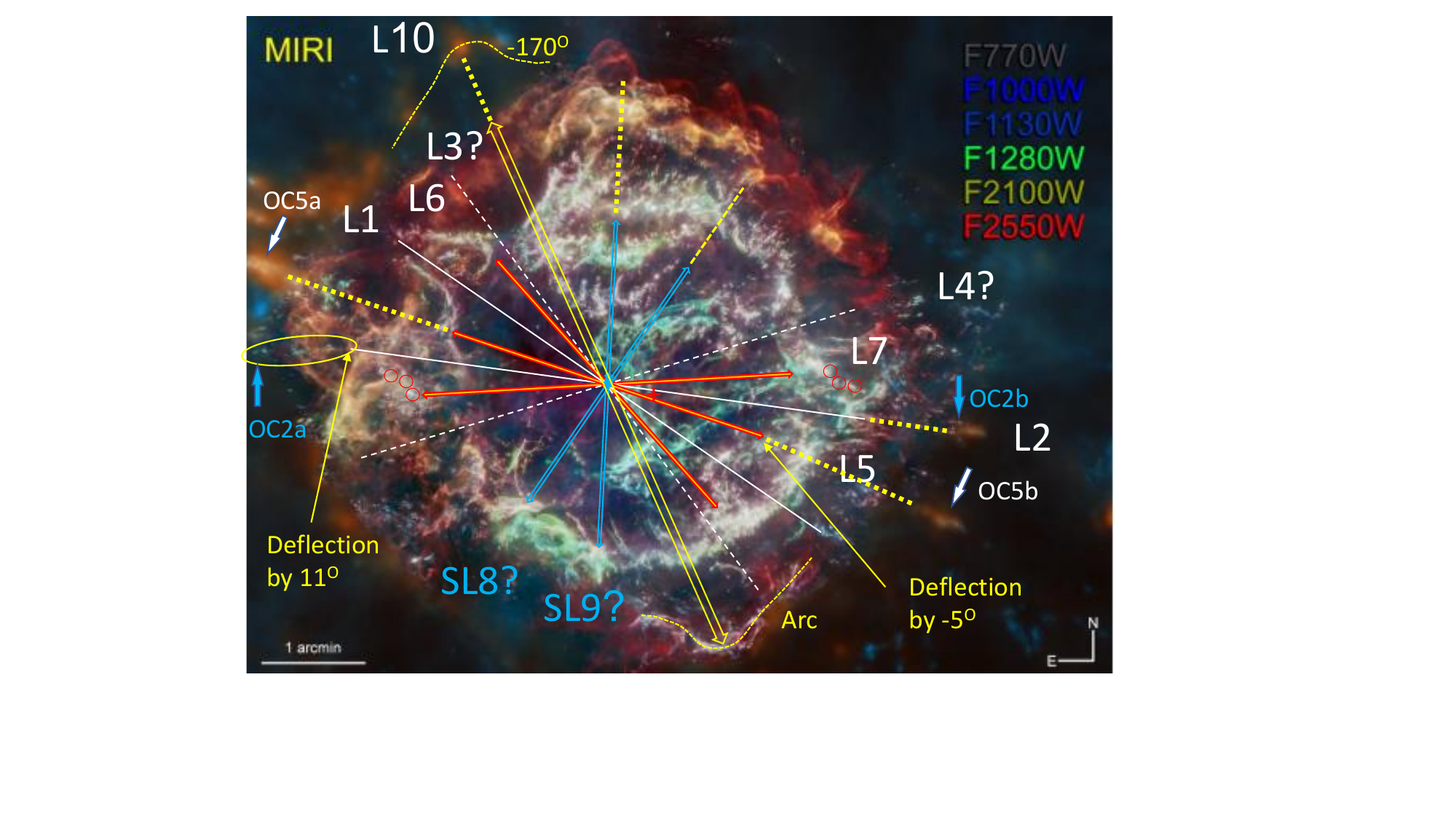}
\end{center}
\caption{A JWST image of Cassiopeia A adapted from \cite{Milisavljevicetal2024}, with our identifications of point-symmetrical morphological futures. On the upper panel, we added our identification of the north protrusion and four outer clumps that we point at with the white or light-blue arrows. The distances of arrow OC2a and arrow OC2b from the cross point of the lines on the lower panel are equal to each other, and so are the distances of arrow OC5a and arrow OC5b.  
In the lower panel, we show that these two pairs of outer clumps are on the continuation of symmetry lines L2 and L5, respectively. The continuation towards OC2a requires a deflection by $11^\circ$ (counterclockwise) and that towards OC5b by $-5^\circ$ (clockwise). We also note that the speculative symmetry axis SL9 points at the north protrusion (upper dashed-yellow line), the largest of 4 protrusions in that region. The JWST reveals another symmetry axis that we mark by a double-line, double-headed yellow arrow with its center at the cross point of the other symmetry axes; we term it L10. It is included now in the point-symmetric wind-rose. The dashed-yellow line marks the arc at the south. We copied it to the north of the remnant and rotated it around itself by $-170 ^\circ$ ($170^\circ$ clockwise).      
}
\label{Fig:JWST} 
\end{figure}

In Figure \ref{Fig:JWST} we present an IR JWST image adapted from \cite{Milisavljevicetal2024}. We mark five structural components on the upper panel, four outer clumps, and a north protrusion. 
In that image, the north protrusion is the largest `ear' of four in the north rim of Cassiopeia A. In the lower panel, we placed the point-symmetric wind-rose from the argon image (Figure \ref{Fig:ArgonMaps1}) on the JWST image. 

Some structural features that define the wind-rose on Figure \ref{Fig:ArgonMaps1} do not appear on Figure \ref{Fig:JWST}, like several clumps. On the other hand, some new clumps and filaments strengthen the point-symmetric wind-rose. Some of the features appearing in Figure \ref{Fig:ArgonMaps1} are related to the presence of argon. This, we suggest, comes from that the jets are active during the explosion process when nucleosynthesis takes place. Therefore, some jet-shaped features are seen in some newly synthesized elements, in this case, argon; they might be less prominent in other wavelengths that are less sensitive to the composition. Some jet are not related to the nucleosynthesis of argon, but shape the ejecta that appears in other wavelengths, here in the IR. The dependence of some morphological features on emission bands is seen in other point-symmetrical morphologies, e.g., the point-symmetric structure of the Cygnus Loop \citep{Shishkinetal2024}.    
The north protrusion is along the northern direction of line SL9. The outer clump OC2a is a filament, a continuation of line L2 on the east with an $11^\circ$ deflection, and the outer clump OC2b is along the western end of line L2. Outer clump OC5a is on the western continuation of line L5, and the outer clump OC5b is along the western continuation of line L5 with a deflection of $-5^\circ$. As we commented before, deflection (bending) commonly occurs in planetary nebulae \citep{SokerHadar2002} and in jets in cooling flows in galaxy clusters and groups (see \citealt{Soker2024CFs} for comparing these deflections of point-symmetric CCSNRs to cooling flows). 

We identify another symmetry axis on the JWST image that we mark as L10 on the lower panel of Figure \ref{Fig:JWST}. It connects an arc on the south, also defined by \cite{Vinketal2024} on an X-ray image, and a bright clump on the north. The yellow double-lined double-headed arrow with its center at the cross point of the other lines marks symmetry axes L10. We drew a line through the southern arc, the dashed-yellow line at the bottom of the lower panel of Figure \ref{Fig:JWST}. We copied this drawn arc to the other side, i.e., north of the remnant, and rotated it around itself by $-170^\circ$  ($170^\circ$ clockwise). This line goes through the three orange clumps in that region.   

\begin{figure}
\begin{center}
\includegraphics[trim=10.3cm 3.3cm 1.9cm 0.0cm ,clip, scale=0.43]{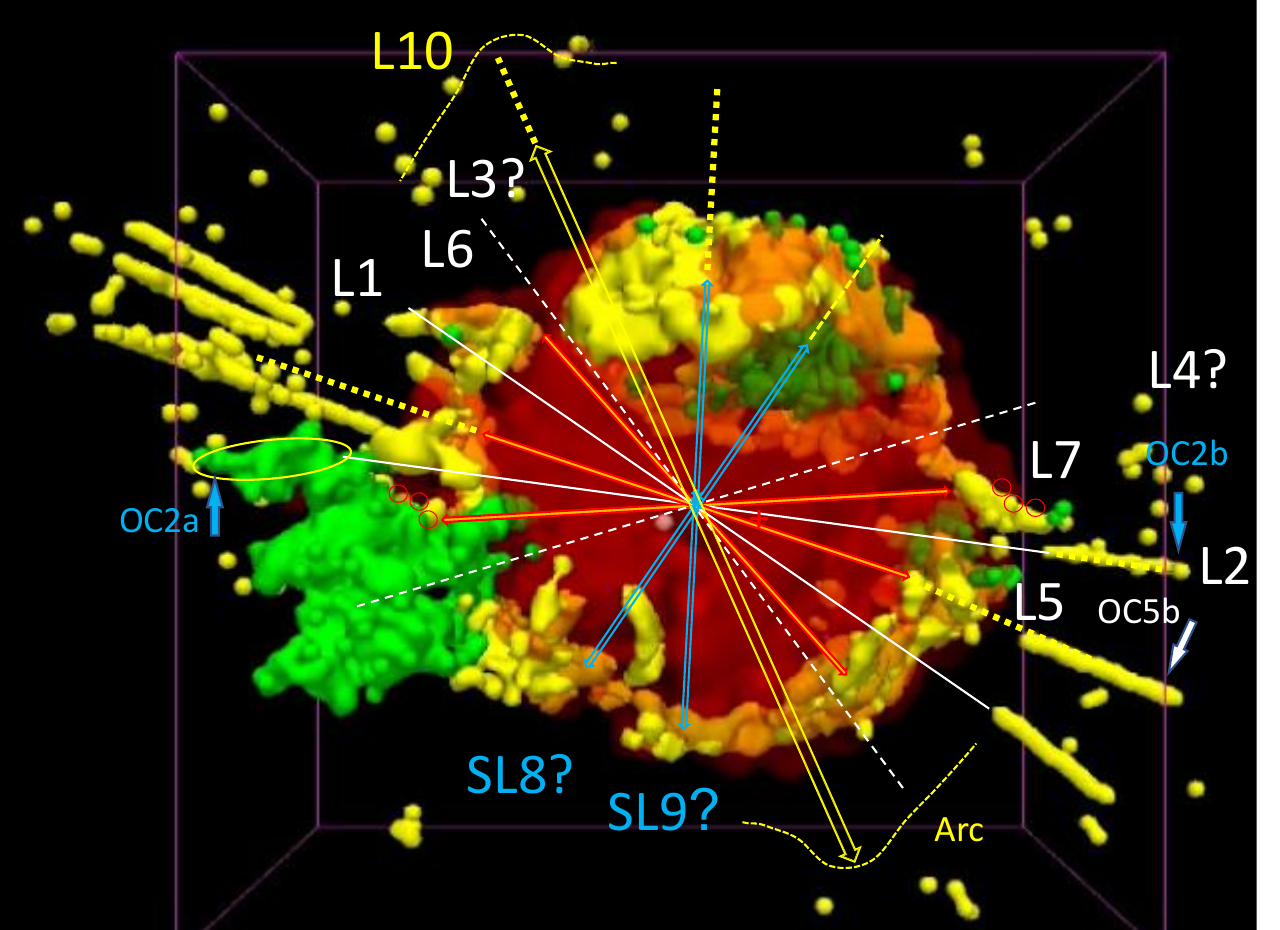}
\end{center}
\caption{A composite image of a three-dimensional reconstruction of Cassiopeia A based on \cite{DeLaneyetal2010}. 
The green and yellow regions are mostly iron (green) in X-rays, the yellow region represents mainly argon and silicon seen in X-rays, optical, and infrared, and the red region is cooler debris seen in the infrared.
We placed the point-symmetric wind-rose that we constructed according to the argon map in Figure \ref{Fig:ArgonMaps1}. 
}
\label{Fig:Visible_animation} 
\end{figure}

Clumps OC2b and OC5b were previously identified in the optical (e.g., \citealt{HammellFesen2008}) and were part of the reconstructed three-dimensional morphology of Cassiopeia A (e.g., \citealt{DeLaneyetal2010, MilisavljevicFesen2013}). In Figure \ref{Fig:Visible_animation} we place the point-symmetric wind-rose on an image based on \cite{DeLaneyetal2010}\footnote{\tiny{ https://www.universetoday.com/23201/cassiopeia-a-comes-alive-in-3-d-movie/}}. The three yellow-colored filaments/jets in the west (right of the figure), ExL1, ExL2, and ExL5, are extensions of lines L1, L2, and L5, respectively. Note that lines L1 and L2 are drawn through the ears and funnels on Figure \ref{Fig:ArgonMaps1}; therefore, their exact directions are uncertain by a few degrees. Considering this uncertainty, the three yellow-colored filaments are along these point-symmetry lines, and clumps OC2b and OC5b in Figure \ref{Fig:JWST} are a part of two of these filaments. The three green-colored iron-rich ejecta fingers on the southwest are about along the symmetry lines L2, L4, and L7, with deflections by several degrees (see Figure \ref{Fig:JWST}). Overall, the image in Figure \ref{Fig:Visible_animation} further strengthens the point-symmetric wind-rose of Cassiopeia A that we build in this study.  

\begin{figure}
\begin{center}
\includegraphics[trim=6.5cm 0.3cm 1.6cm 0.3cm ,clip, scale=0.41]{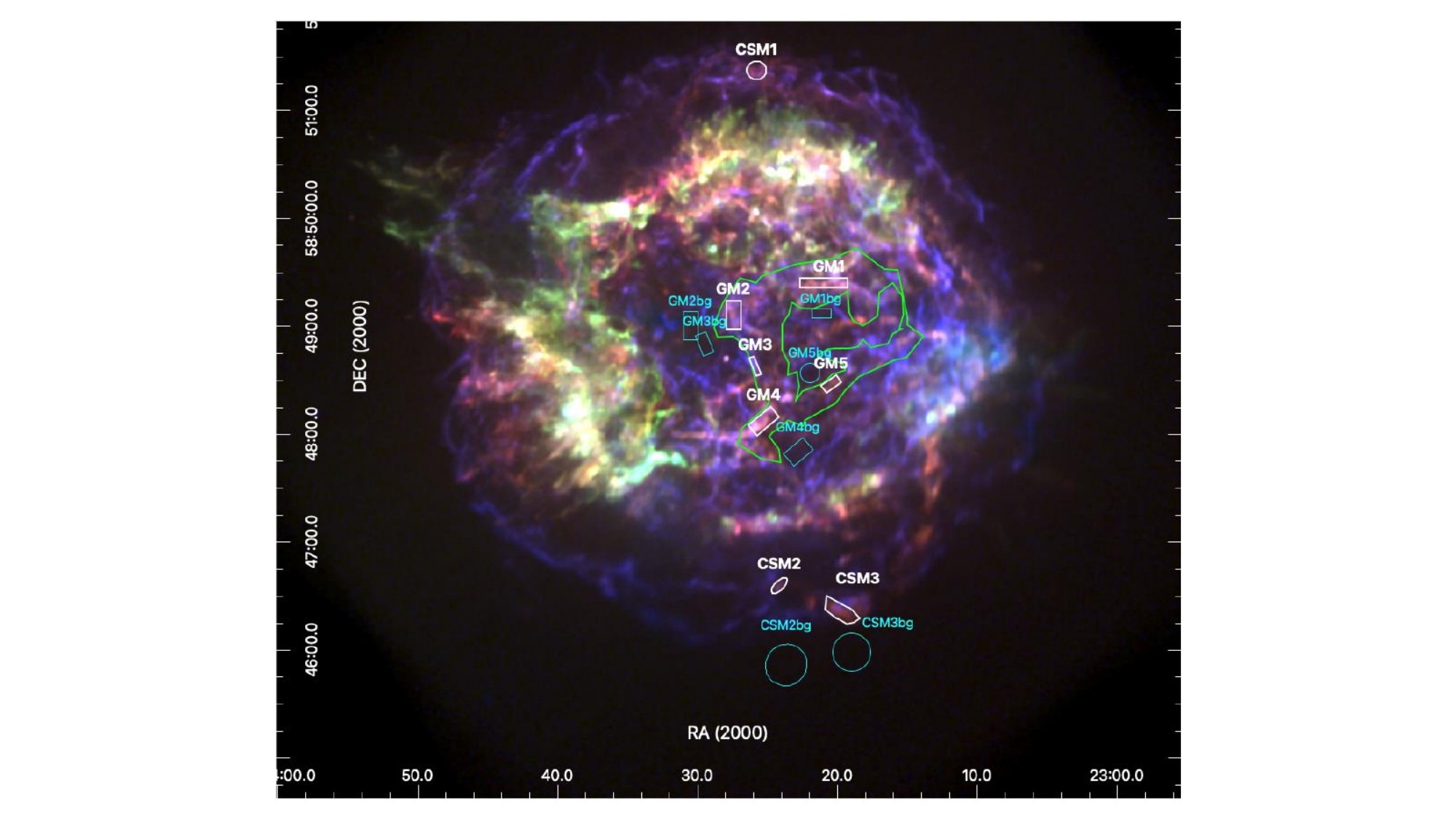}
\includegraphics[trim=6.5cm 0.3cm 1.6cm 0.3cm ,clip, scale=0.39]{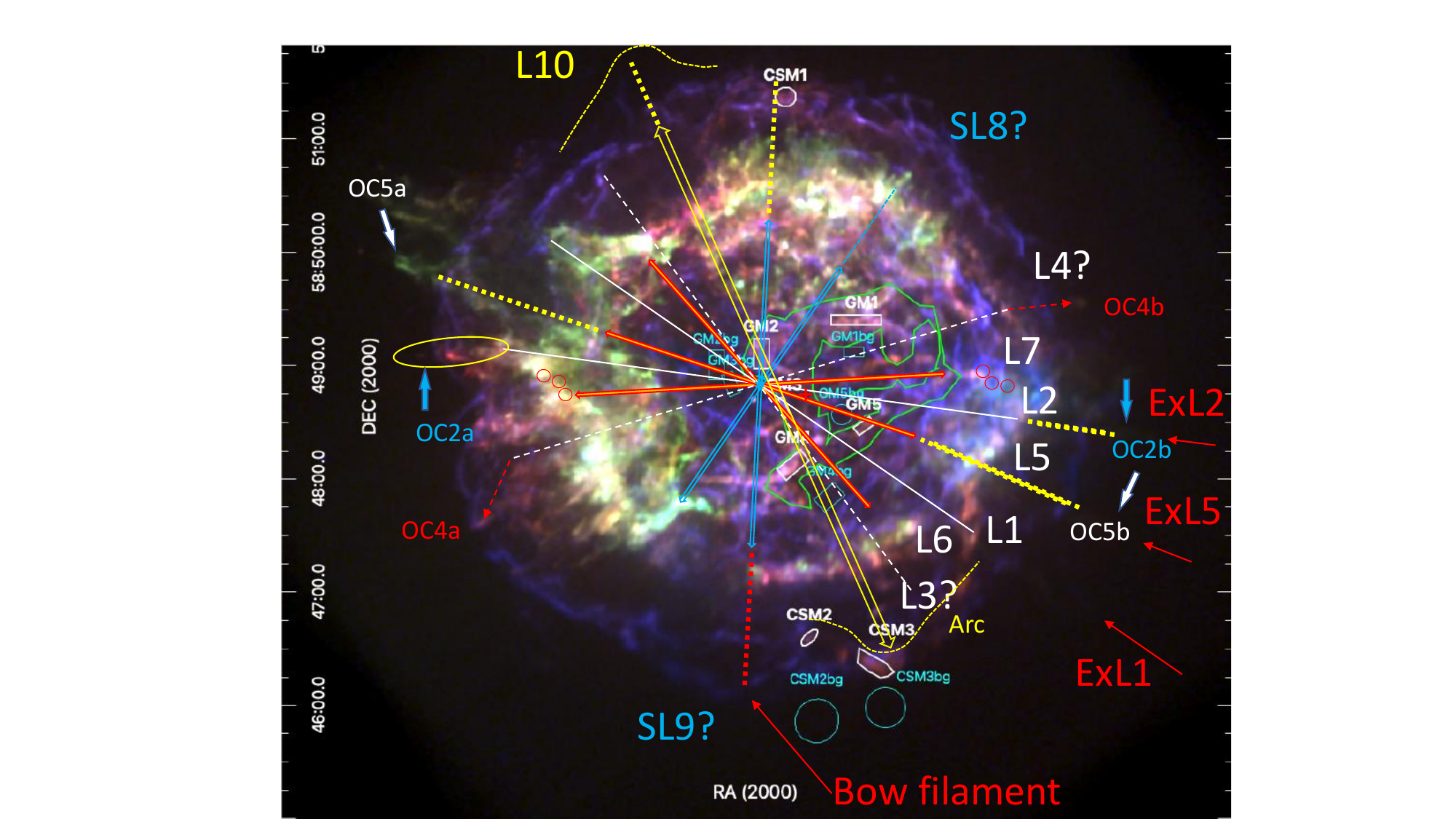}
\end{center}
\caption{An X-ray Chandra RGB color image of Cassiopeia A adapted from \cite{Vinketal2024}; the two panels have the same X-ray image. Red represents 1.2–1.4 keV  (showing Fe-L and Mg XI line emission),
green represents 1.85-1.98 keV (Si XII), and blue represents 4.0–4.5 keV (continuum). Green polygons border the `Green Monster' structure. 
In the lower panel, we placed the point-symmetric wind-rose from Figure \ref{Fig:ArgonMaps1} and added some labels. The filament extensions of three symmetry lines that are the same as in Figure \ref{Fig:Visible_animation}, are marked by ExL1, ExL2, and ExL5, respectively. Here, they are seen as faint purple-colored filaments, as the bow filament is. 
We also mark two clumps that might be connected to symmetry line L4; these need further study.  
}
\label{Fig:NewXray} 
\end{figure}

\cite{Vinketal2024} present new X-ray images of Cassiopeia A. We bring one image in Figure \ref{Fig:NewXray}, in the upper panel as presented by \cite{Vinketal2024} and in the lower panel with our point-symmetric wind-rose and some additional marks. There are three very faint purple-colored filaments corresponding to the yellow-colored southwest filaments in Figure \ref{Fig:Visible_animation}, and a bow-shaped filament along the continuation of symmetry line SL9. The new X-ray image does not introduce new symmetry lines but further strengthens our constructed point-symmetric wind-rose. 

 As with other wavelengths, we treat the X-ray images in the different X-ray bands as they appear in the cited observations, and do not go into analyzing the processes that determine the X-ray emission. The X-ray emission comes from gas that the reverse shock heated. The emission seem to amplify inhomogeneities. For example, the X-ray knots (clumps) are concentrations of heavy elements \citep{LamingHwang2003}. Instabilities that developed after the passage of the reverse shock can destroy such clumps, hence the appearance might change over time. In this study we do not consider these processes which will have to be included in future numerical simulations of the shaping by jets.

Our study of Cassiopeia images concludes that there is a robust point-symmetric morphology on the plane of the sky described by the point-symmetric wind-rose we constructed.  If most of the ejecta of Cassiopeia A are in a ring (or torus) structure as suggested by earlier studies (e.g., \citealt{Willingaleetal2003, DeLaneyetal2010, MilisavljevicFesen2013}), namely, in a plane, there is another symmetry axis, one through the center and perpendicular to the plane of the ring (close to being along the line of sight).  However, it is not clear yet whether the mass is indeed concentrated in such a plane. The inhomogeneous medium into which the inhomogeneous ejecta expand and the ejecta itself imply many dense zones; the densest zones emit the most but do not necessarily follow most of the mass. If the mass is indeed in a plane, then most of the jittering jets that shaped the ejecta were active in the plane of the ring. The jets' jittering (section \ref{sec:intro}) and the jittering in a plane, planar jittering, are as expected in the JJEM. 

\section{Summary} 
\label{sec:Summary}
 
We analyzed the morphology of the CCSNR Cassiopeia A and identified a point-symmetric structure shaped by at least two, but more likely more than four, pairs of jets. We first identified point symmetry in the argon X-ray image from \cite{DeLaneyetal2010}, and built a \textit{point-symmetric wind-rose} composed of seven symmetrical lines (axes) connecting pairs of opposite structural features, lines L1-L7 (Figure \ref{Fig:ArgonMaps1}). We also speculated on two other lines, SL8 and SL9. Placing this point-symmetric wind-rose on the newly released JWST image \citep{Milisavljevicetal2024}, on an image based on \cite{DeLaneyetal2010}, and on a new X-ray image  \citep{Vinketal2024}, in Figures \ref{Fig:JWST} - \ref{Fig:NewXray}, respectively, we strengthened the claim for a point symmetry. 
From the new JWST image (Figure \ref{Fig:JWST}) we identified another symmetry axis, L10. 

Not all lines indicate pairs of jets. Some dense clumps might be formed by the compression of ejecta between two jet-inflated lobes/bubbles (e.g., \citealt{Soker2024CFs}). The exact relation between jets' properties and geometry and the morphology they form requires three-dimensional simulations.  

The robust point symmetry of Cassiopeia A is as expected in the JJEM but has no explanation in the competing delayed neutrino explosion mechanism; for example, post-explosion jets cannot explain most point-symmetric morphological properties of CCSNRs \citep{SokerShishkin2024}.  Moreover, the jittering jets might be in more or less the same plane (termed planar jittering), which is also an expectation of the JJEM in some CCSNe. Indeed, \cite{PapishSoker2014Planar} already speculated that the torus morphology of a tilted thick disc with multiple jets in Cassiopeia A, as observations find (e.g., \citealt{Willingaleetal2003, DeLaneyetal2010, MilisavljevicFesen2013}), is a result of a planar jittering pattern. We confirmed their speculation with our new identification of the point symmetry of Cassiopeia A. 

The reason for a tendency for planar jittering is as follows \citep{PapishSoker2014Planar}. The jets expel infalling core material in their vicinity. This leaves material to be accreted along a direction perpendicular to the jets' axis. The action of two jet launching episodes that do not share the same axes is to allow accretion along a direction perpendicular to the plane the two jets' axes define. This accreted gas has angular momentum in the plane defined by the previous two axes, perpendicular to its inflow direction. Therefore, the new axis of the pair of jets will be in the same plane defined by the first two jets' axes. This causes the tendency for planar jittering. Large angular momentum fluctuations of the accreted gas prevent jittering in a thin plane and, in many CCSNe, are sufficiently large to prevent the planar jittering altogether. 

The neutrino-driven explosion mechanism also has difficulties in explaining the elemental distribution in Cassiopeia A. \cite{HwangLaming2012} and \cite{LamingTemim2020} find most of the iron to be in the outer ejecta. They argued that this and the mismatch of iron and titanium distributions are incompatible with the prediction of the neutrino-driven explosion mechanism (as also \citealt{Soker2017RAA} argued). A much more vigorous overturn of the stratification has happened than expected by convective or Rayleigh-Taylor instabilities in the neutrino-driven explosion mechanism. We speculate here that jets can supply this vigorous overturn.

\cite{Satoetal2023} calculate the production of $^{55}$Mn in CCSNe with and without neutrino-ejecta interaction and compared to the measured value of Mn/Fe in Cassiopeia A. They conclude that the Mn/Fe ratio is a robust indication of neutrino interaction with the inner ejecta. The neutrino flux in the JJEM is very similar to the neutrino-driven explosion mechanism, as the neutrino flux mainly results from the cooling of the newly born NS (which is the same in both explosion mechanisms). Therefore, the JJEM expected production of $^{55}$Mn is very similar to the calculation by \cite{Satoetal2023}. 
Another point to emphasize is that neutrino heating does boost the explosion in the JJEM \citep{Soker2022Boosting}, but neutrino heating does not play the primary role in driving the explosion. We argue that the point-symmetric morphologies our research group identified in about ten CCSNRs testify to the primary role of jittering jets. Overall, it is difficult to use the outcomes of neutrino interaction with the ejecta to distinguish between the neutrino-driven explosion mechanism and the JJEM. The morphologies of many CCSNRs show the clearest distinguishing with a decisive favoring of the JJEM.

The point symmetry of Cassiopeia A is the richest in the diverse types of morphological features, and the number of symmetry axes among the nine CCSNRs with claimed point symmetry (Cassiopeia A; Vela SNR; SNR N63A; SNR G321.3–3.9; SNR 1987A; SNR CTB 1; SNR G107.7-5.1; SNR 0540-69.3; see section \ref{sec:intro} for references and \citealt{Soker2024CFs} for the studies of these eight SNRs concerning jet shaping in cooling flows;  for the Cygnus Loop see  \citealt{Shishkinetal2024}).   As such, our study substantially adds to the accumulating observational support of the JJEM as the primary explosion mechanism of CCSNe.  

\section*{Acknowledgements}
We thank the referee for constructive comments.  This research was supported by a grant from the Israel Science Foundation (769/20).

\end{document}